\documentclass[a4paper,11pt]{article}
\usepackage{jcappub} 
\usepackage{lineno}
\usepackage{graphicx,times}             
\usepackage{natbib}
\usepackage{amssymb,amsmath}

\arxivnumber{} 
\title{\boldmath Halo Spin Depends on The Distance to Cosmic Filament}

\author{Shihong Liu$^{1,2}$, Yu Rong$^{1,2,*}$}
\affiliation{1.Department of Astronomy, University of Science and Technology of China, Hefei, Anhui 230026, China}
\affiliation{2.School of Astronomy and Space Sciences, University of Science and Technology of China, Hefei 230026, Anhui, China}

\emailAdd{rongyua@ustc.edu.cn}

\abstract{We employ a semi-analytical methodology to estimate the dark matter halo spin of HI-rich galaxies in the Arecibo Legacy Fast Alfa Survey and investigate the relationship between halo spin and the proximity of galaxies to cosmic filaments. We exclude galaxies with low HI signal-to-noise ratios, those potentially influenced by velocity dispersions, and those affiliated with galaxy clusters/groups. Additionally, we apply a mass-weighting technique to ensure consistent mass distribution across galaxy samples at varying distances from filaments. Our analysis reveals, for the first time, a subtle yet statistically significant correlation between halo spin and filament distance in observational data, indicating higher spins closer to filaments. This suggests that the tidal forces exerted by filaments may impact the spin of dark matter halos.}

\keywords{galaxies: formation --- galaxies: evolution --- methods: statistical}

\begin{document}
\maketitle
\flushbottom

\section{Introduction}
\label{sec:1}

The spin of dark matter halos plays a crucial role in shaping the evolution and characteristics of galaxies. Hydrodynamical simulations \cite[e.g.,][]{Kim13,Jiang192} and semi-analytic galaxy formation models \cite[e.g.,][]{Mo98} suggest that halo spin significantly influences the size and density of baryonic matter distribution, particularly in massive late-type galaxies. While the impact of halo spin on low-mass galaxies remains a topic of debate \citep[e.g.,][]{Yang23}, studies on specific dwarf galaxies, such as ultra-diffuse galaxies \cite{Rong17a,Amorisco16,Liao19,Benavides23}, indicate that halo spin may have a notable effect on the distribution of baryonic matter in dwarf galaxies. Therefore, understanding the influence of halo spin on baryonic matter within galaxies and their co-evolution is crucial for advancing our understanding of galaxy formation models.

Moreover, in addition to its connection with internal galaxy properties, halo spin is also linked to the large-scale structure of the universe. Halo spin is thought to originate from tidal torques induced by large-scale structures, resulting from gravitational interactions with neighboring structures \cite[e.g.,][]{Peebles69,White84} or through mergers \cite[e.g.,][]{Gardner01,Vitvitska02,Hetznecker06,Maller02}. Simulation studies suggest that halos exhibit faster spins in stronger tidal fields, with a more pronounced effect observed in more massive halos \cite{Wang11}. As halos transition from the linear to non-linear phases towards virialization, the impact of tidal torque diminishes. During this transition, the flow field surrounding halos displays non-zero vorticity, which is critical in determining halo angular momentum and leading to alignment between halo spin and vorticity \cite[e.g.,][]{Liberskind13}. Theoretically, a correlation between halo spin and large-scale filaments should exist due to the strong link between vorticity and filament direction. However, observational evidence suggests that the correlation between halo spin and filaments, particularly in late-type galaxies, is minimal and, in some instances, almost non-existent. The current understanding of the relationship between halo spin and the environment remains uncertain, with conflicting results from observations and simulations. Understanding dark matter halo spin and its relationship with large-scale structures is essential for unraveling the formation and evolution of galaxies in the universe.

In this study, we utilize a semi-analytic approach to estimate halo spin for a substantial sample of HI-bearing galaxies from an HI survey and investigate the dependence of halo spin on galaxy distances to large-scale filaments. Section~\ref{sec:2} presents the dataset and describes the methodology for estimating halo spin. Section~\ref{sec:3} provides a statistical analysis of galaxy halo spins across different filament distance ranges. Our findings are summarized in section~\ref{sec:4}.

\section{Data}
\label{sec:2}

\subsection{Sample}

The selection of galaxies for this study is based on the extensive Arecibo Legacy Fast Alfa Survey \citep[ALFALFA;][]{Giovanelli05,Haynes18}, a wide-ranging extragalactic HI survey covering approximately 6,600 deg$^2$ at high Galactic latitudes. A comprehensive catalog \citep[$\alpha.$100;][]{Haynes18} released by the ALFALFA collaboration includes about 31,500 sources with radial velocities below 18,000 km s$^{-1}$, providing essential properties such as signal-to-noise ratio (SNR) of the HI spectrum, cosmological distance, 50\% peak width of the HI line ($W_{50}$) corrected for instrumental broadening, and the HI mass ($M_{\rm{HI}}$), among other parameters. Further details on these properties and their uncertainties can be found in \cite{Haynes18}.

To augment the dataset, ALFALFA galaxies have been cross-referenced with SDSS data \citep{Alam15}. Previous investigations by \cite{Durbala20} have estimated stellar masses $M_{\star}$ for ALFALFA galaxies with optical counterparts using different methods, with a preference for the stellar mass derived from spectral energy distribution (SED) fitting. In cases where UV or infrared data are lacking for SED fitting, the stellar mass based on $g-i$ color is utilized, with any discrepancies in stellar mass among these methods considered negligible.

\subsection{Rotation velocity}

The rotation velocity is computed as $V_{\rm{rot}}=W_{50}/2/\sin\phi$, where $\phi$ denotes the inclination of the HI disk. When resolved HI data is unavailable, the optical apparent axis ratio $b/a$ from \cite{Durbala20} is employed to estimate the HI disk inclination $\phi$. The calculation involves $\sin\phi=\sqrt{(1-(b/a)^2)/(1-q_0^2)}$ (if $b/a \leq q_0$, we set $\phi=90^{\circ}$), with a fixed intrinsic thickness $q_0\sim 0.2$ \citep{Tully09,Giovanelli97,Li21} for massive galaxies, and $q_0\sim 0.4$ \cite{Rong24} for low-mass galaxies with $M_{\star}<10^{9.5}\ M_{\odot}$.

To ensure accuracy, galaxies with inclinations $\phi<50^{\circ}$ and low SNR$<10$ are excluded from the analysis. Additionally, galaxies exhibiting `single-horned' HI line profiles, indicating dominance of velocity dispersion over regular rotation, are identified using the kurtosis parameter $k_4>-1.0$ following \cite{Hua24} and \cite{ElBadry18}. Only galaxies with double-horned HI profiles are considered in this study, excluding dispersion-dominated systems.

The final sample comprises approximately $3,280$ galaxies, with stellar masses ranging from $10^7$ to $10^{11}$ $\rm M_{\odot}$, as depicted in panel~a of Fig.~\ref{fig1}. 

\subsection{Halo spin}

Under the assumption of an isothermal sphere model for the galaxy's dark matter halo and neglecting baryonic matter effects, the halo spin ($\lambda_{\rm{h}}$) can be expressed as
$\lambda_{\rm{h}}\sim R_{\rm{HI,d}}/V_{\rm{rot}}^{3/2}$ \cite{Hernandez07},
where $V_{\rm{rot}}$ represents the halo's rotation velocity.
The scale length of the HI disk, $R_{\rm{HI,d}}$, is determined assuming a thin gas disk in centrifugal balance \citep{Mo98}, characterized by an exponential surface density profile:
\begin{equation}
	\Sigma_{\rm{HI}}(R)=\Sigma_{{\rm{HI}},0} {\rm{exp}}(-R/R_{{\rm{HI,d}}}),
\end{equation}
where $\Sigma_{{\rm{HI}},0}$ denotes the central surface density of the HI disk. The total HI mass $M_{\rm{HI}}$ is linked to the scale length by
\begin{equation} M_{\rm{HI}} = 2 \pi \Sigma_{{\rm{HI}},0} R_{{\rm{HI,d}}}^2 \label{HIeq_mass}. \end{equation} 
Furthermore, we define the HI radius $r_{\rm{HI}}$ as the radius at which the HI surface density reaches $1\ \rm M_{\odot}\rm{pc^{-2}}$. The estimation of $r_{\rm{HI}}$ is guided by the observed correlation between $r_{\rm{HI}}$ and $M_{\rm{HI}}$, as indicated by empirical studies: $\log r_{\rm{HI}}=0.51\log M_{\rm{HI}}-3.59$ \citep{Wang16,Gault21}. Hence, at $r_{\rm{HI}}$, we have 
\begin{equation} \Sigma_{{\rm{HI}},0} {\rm{exp}}(-r_{\rm{HI}}/R_{{\rm{HI,d}}})=1\ \rm  M_{\odot}\rm{pc^{-2}}. \label{HIeq_3} \end{equation} 
Using equations~(\ref{HIeq_mass}) and (\ref{HIeq_3}), we can determine the value of $R_{{\rm{HI,d}}}$ for each galaxy in our sample, thereby enabling the estimation of the halo spin.

\subsection{Distance to filament}

For each galaxy, we identify its corresponding large-scale filament using the filament catalog compiled by \cite{Tempel14a}. 
This catalog contains the distance of each SDSS galaxy to the nearest filament spine, denoted as $d_{\rm{gf}}$, which we utilize to explore the relationship between halo spin and filament proximity. The catalog also provides the richness $N$ of the galaxy group to which each galaxy belongs. In our analysis, we focus on galaxies with $N=1$, selecting isolated HI-bearing galaxies to investigate their halo spins.

To further refine our analysis, we divide our sample of HI-bearing galaxies into subgroups based on their proximity to filaments:  $d_{\rm{gf}}\leq 1.0$~Mpc$/h$, $1.0<d_{\rm{gf}}\leq 3.0$~Mpc$/h$, and $d_{\rm{gf}}> 3.0$~Mpc$/h$. Typically, a distance of $d_{\rm{gf}}\leq 1$~Mpc$/h$ corresponds to the approximate `boundary' of a filament  \citep{Wang24}.


\section{Results}
\label{sec:3}

\begin{figure}
   \centering
   \includegraphics[width=\textwidth, angle=0]{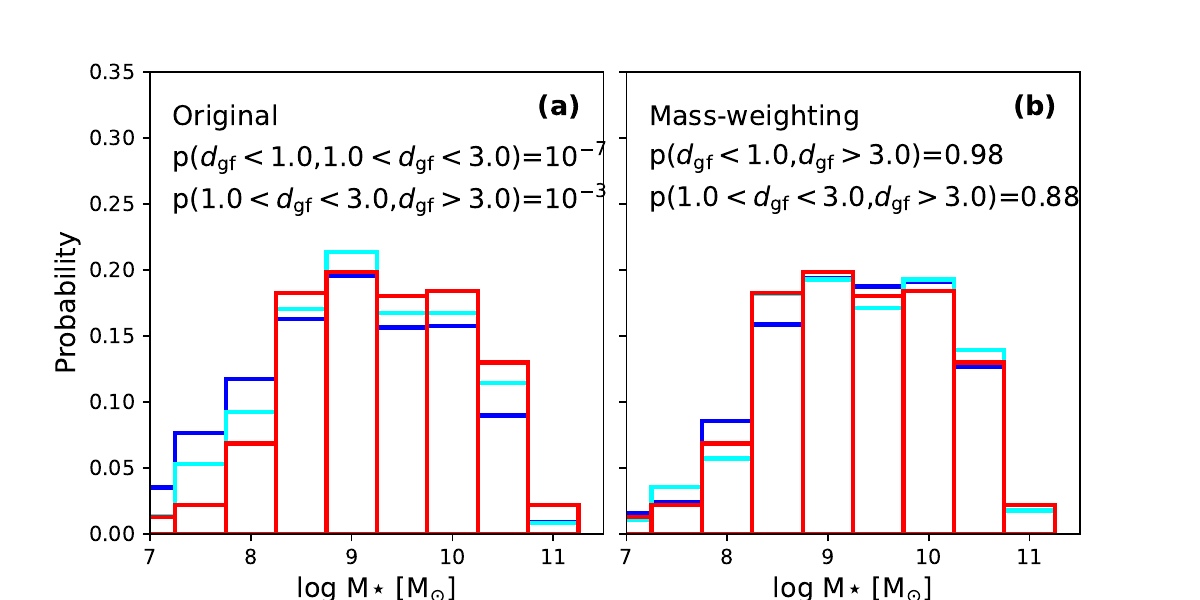}
   \caption{The distributions of stellar masses of the galaxy subsamples with $d_{\rm{gf}}\leq 1.0$~Mpc$/h$ (red), $1.0<d_{\rm{gf}}\leq 3.0$~Mpc$/h$ (cyan), and $d_{\rm{gf}}> 3.0$~Mpc$/h$ (blue). Panels~a and b correspond to the original galaxy samples and samples after mass-weighting method, respectively. }
   \label{fig1}
   \end{figure}

The stellar mass distributions of the three subsamples are presented in panel~a of Fig.~\ref{fig1}, showing distinct differences supported by low $p$-values from Kolmogorov-Smirnov (K-S) tests. Given the known relationship between halo spin and galaxy mass, it is crucial to ensure that the stellar mass distributions across the three subsamples are comparable to investigate the impact of environment on halo spins. To achieve this, we employ a mass-weighting control method.

Initially, we establish the stellar mass distribution of the $d_{\rm{gf}}\leq 1.0$~Mpc$/h$ subsample as the reference. Subsequently, we assign weights to the other subsamples using the equation:
\begin{equation}
    w_{x}(m_{\star})=f_{d_{\rm{gf}}\leq 1.0}(m_{\star})/f_{x}(m_{\star}),
\end{equation}
where $w_{x}(m_{\star})$ and $f_{x}(m_{\star})$ represent the weights and fraction of galaxies in subsample $x$ with stellar masses between $m_{\star}$ and $m_{\star}+{\rm d}m_{\star}$, respectively. We then randomly select $n(m_{\star})$ galaxies from the subsample $x$ within the stellar mass range $m_{\star}$ to $m_{\star}+{\rm d}m_{\star}$ using
\begin{equation}
    n(m_{\star})=w_{x}(m_{\star})/{\rm max}(w_{x}(m_{\star}))*n_x(m_{\star}),
\end{equation}
where $n_x(m_{\star})$ denotes the number of galaxies in subsample $x$ within the stellar mass interval $m_{\star}$ to $m_{\star}+{\rm d}m_{\star}$, and ${\rm max}(w_{x}(m_{\star}))$ is the maximum value of $w_{x}(m_{\star})$ across different mass bins. The stellar mass distributions of the three subsamples post mass-weighting are compared in panel~b of Fig.~\ref{fig1}. Subsequently, we utilize the mass-weighted subsamples to examine the relationship between halo spin and $d_{\rm{gf}}$.

\begin{figure}
   \centering
   \includegraphics[width=\textwidth, angle=0]{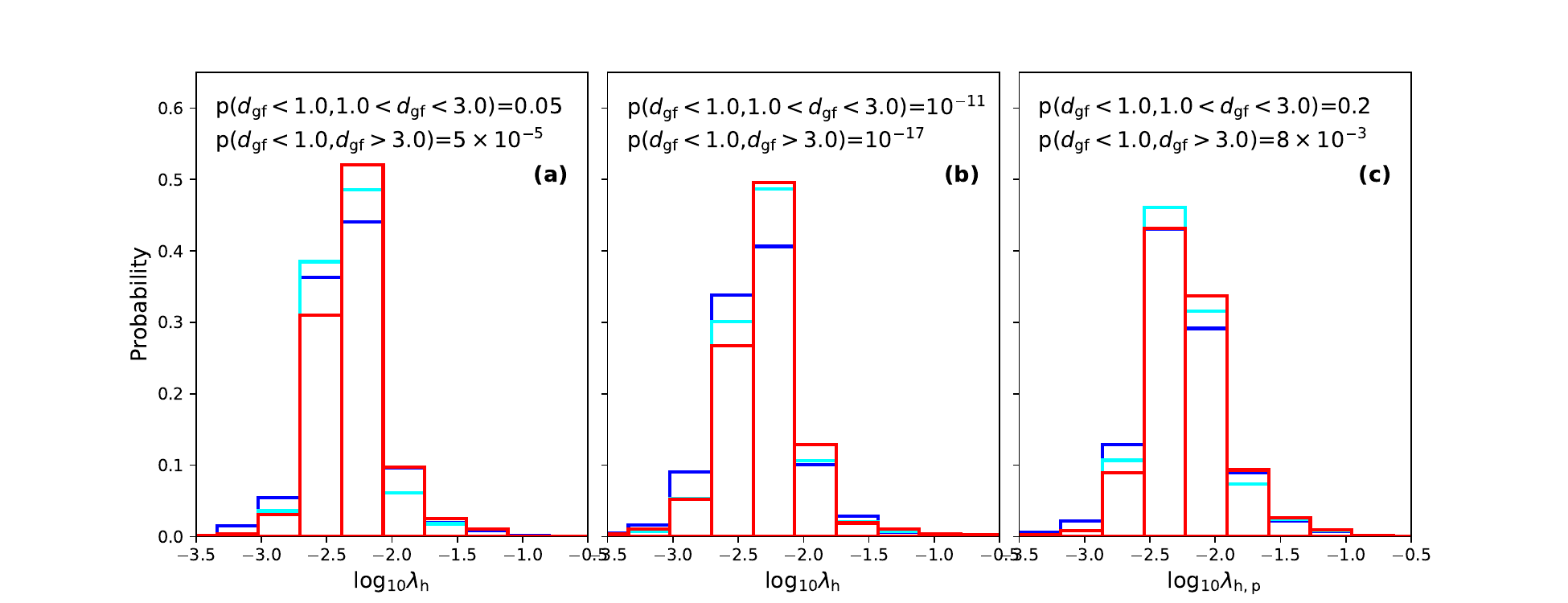}
\caption{Comparison of the halo spin distributions for subsamples with $d_{\rm{gf}}\leq 1.0$~Mpc$/h$ (red), $1.0<d_{\rm{gf}}\leq 3.0$~Mpc$/h$ (cyan), and $d_{\rm{gf}}> 3.0$~Mpc$/h$ (blue). Panel (a) shows results using optical inclinations; panel (b) presents results after accounting for inclination uncertainties via Monte Carlo sampling of misaligned HI disk orientations; panel (c) displays the distributions derived using the practical halo spin $\lambda_{\rm h,p}=\lambda_{\rm h}/f_{\lambda}$, where $\lambda_{\rm h}$ is the HI-derived halo spin from our semi-analytical method and $f_{\lambda}$ follows a Gaussian distribution with mean 1.27 and standard deviation 0.40, truncated to ensure positivity. The small $p$-values from two-sample Kolmogorov–Smirnov tests indicate statistically significant differences among the spin distributions in different filament environments. }
   \label{compare}
   \end{figure}

In Fig.~\ref{compare}, we present a comparative analysis of halo spin distributions across three distinct subsamples. Our results reveal a subtle yet statistically significant correlation between the spin properties of galaxy halos and their proximity to large-scale filaments. Specifically, the median spins $\log\lambda_{\rm{h}}$ are determined as $-2.16\pm0.01, -2.19\pm0.01, -2.20\pm0.01$ for the subsamples characterized by $d_{\rm{gf}}\leq 1.0$~Mpc$/h$, $1.0<d_{\rm{gf}}\leq 3.0$~Mpc$/h$, and $d_{\rm{gf}}> 3.0$~Mpc$/h$, respectively. The uncertainties in the median values are derived as $\sigma_{{\lambda}}/\sqrt{N}$, where $N$ represents the sample size and $\sigma_{{\lambda}}$ signifies the standard deviation. Notably, the disparity in median $\log\lambda_{\rm{h}}$ between the subsamples of  $d_{\rm{gf}}\leq 1.0$~Mpc$/h$ and $d_{\rm{gf}}>3$~Mpc$/h$ is statistically significant at a confidence level of approximately $4\sigma$. The two-sample K-S tests also yield small $p$-values further corroborating the observed distinctions in halo spin distributions among the three subsamples, as depicted in panel~a of Fig.~\ref{compare}. Our findings suggest a discernible association between halo spin and the proximity to filaments, particularly highlighting the substantial divergence in halo spins between galaxies situated within filaments with $d_{\rm{gf}}\leq 1.0$~Mpc$/h$ and those positioned outside filaments. This observation underscores the potential influence of large-scale filament tidal fields on galaxy halo spin. Notably, the trend of increasing halo spins in denser environments aligns with previous studies based on $N$-body simulations \citep{Wang11,Hahn07b}.

\section{Summary and Discussion}
\label{sec:4}

We have employed a semi-analytical method to assess the halo spins of the HI-bearing galaxy sample obtained from ALFALFA and explored the potential correlation between spin values and the proximity of galaxies to large-scale filaments. To enhance the precision of our investigation, we have excluded galaxies within galaxy clusters/groups and those potentially affected by velocity dispersions. Furthermore, we have considered the influence of galaxy masses on the variations in halo spin. Our results indicate a subtle yet statistically significant relationship between halo spins and the distances of galaxies to filaments, implying a potential influence of large-scale filament tidal fields on galaxy halo spin.

It is important to note that the spins of dark matter halos, their internal gas, and stellar components may not be perfectly aligned \citep[e.g.,][]{vandenBosch02,Teklu15,Velliscig15,Bett10,Zjupa17}. 
In this work, the optical inclinations were employed in this study to estimate the rotation velocities and halo spins of galaxies. This method introduces uncertainties due to small misalignments ${\rm d}\phi$ between optical and HI inclinations observed in various galaxies \citep{Hunter12,Oh15} and simulations \citep{Nelson18,Nelson19,Vogelsberger14}. While these uncertainties may slightly attenuate any potential halo spin dependence on filament distance, the majority ($\gtrsim 70\%$) of galaxies exhibit ${\rm d}\phi<20^{\circ}$, indicating a moderate impact of misalignment. Here, we quantitatively estimate the influence of inclination misalignment on the correlation between spin and distance to filament. We use the Monte Carlo method and set the misaligned ${\rm d}\phi$ to follow a Gaussian distribution with a center of 0 and sigma of $20^{\circ}$. Using the reset HI inclination, we re-calculate the halo spin of each galaxy and compare the spin distributions of subsamples with $d_{\rm{gf}}\leq 1.0$~Mpc$/h$, $1.0<d_{\rm{gf}}\leq 3.0$~Mpc$/h$, and $d_{\rm{gf}}> 3.0$~Mpc$/h$. As indicated in panel~b of Fig.~\ref{compare}, the distributions of the different environments are still different, suggesting that the small misalignment would not erase the spin dependence on distance to cosmic filament.

Indeed, cosmological simulations robustly demonstrate a strong correlation between the magnitude of the gas spin and that of its host dark matter halo \citep[e.g.,][]{vandenBosch02,Teklu15,Sharma05}, despite known misalignments among galactic components. For instance, \cite{Sharma05} found that the mean ratio of gas spin to halo spin is approximately 1.27, with a standard deviation of 0.40. Statistically, halos with higher gas spin also exhibit higher halo spin, and vice versa. This correlation justifies our use of HI gas kinematics as a proxy to study the environmental dependence of halo spin. To assess the impact of the scatter in this ratio, we follow \cite{Sharma05} and define a practical halo spin as $\lambda_{\rm h,p}=\lambda_{\rm h}/f_{\lambda}$, where $\lambda_{\rm h}$ is the HI-derived halo spin from our semi-analytical method and $f_{\lambda}$ follows a Gaussian distribution with mean 1.27 and standard deviation 0.40 (truncated to ensure positivity). As illustrated in panel~c of Fig.~\ref{compare}, the dependence of halo spin on distance to the filament spine persists, albeit somewhat weakened. Thus, while the HI-derived spin may be imprecise for individual galaxies, it provides a statistically reliable proxy for comparing relative halo spin differences across environments.

Additionally, it is worth noting that the filamentary network is complex, intricate, and multiscale. The filament catalog we used, based on the Bisous algorithm, primarily identifies filaments of $\sim2$~Mpc thickness but may miss numerous thin tendrils branching off main arms or fail to represent massive $\sim4$~Mpc filaments as single coherent structures. Filaments indeed exhibit hierarchical organization: backbones are more continuous and thicker (typical diameter $\sim 2\--4$~Mpc), whereas outskirts contain short, thin branches often linked to low-mass nodes or isolated galaxies. Studies indicate that backbone filaments reside in higher-density environments, while tendril-like structures occupy lower-density regions \citep[e.g.,][]{Cautun14}. Given our findings and prior theoretical work on tidal fields, low-density environments exert weaker influences on halo spin \citep[e.g.,][]{Wang11,Hahn07b}. We therefore speculate that in such tendrils, halo spin may show little variation with distance from the filament spine, as the ambient density remains low regardless of proximity. Including these unresolved tendrils would likely introduce random noise into our analysis. Conversely, if massive filaments were fragmented by the detection algorithm into multiple spines, our current measurement would already incorporate additional noise\---tending to dilute, rather than fabricate, the observed correlation. Hence, our conclusions are conservative. In fact, true massive filaments possess even higher densities, implying stronger tidal fields near their genuine spines and thus a potentially more pronounced effect on halo spin.

Note that the presented analysis relies on several semi-analytical relations between galaxy properties, which introduce non-negligible uncertainties in halo spin estimates, particularly for individual systems. However, our primary objective is a statistical comparison across a large galaxy sample, not precise spin measurements for single objects. The estimator adopted here follows the widely used formalism of \cite{Hernandez07}, rooted in the classical disk formation model of \cite{Mo98}. These methods have been extensively applied in observational studies to infer statistical halo spin distributions \citep[e.g.,][]{Cervantes-Sodi08,Leisman17,Huang12}, despite known limitations at the individual level. Moreover, halo spin distributions derived from such semi-analytical approaches show good statistical agreement with those from cosmological simulations \citep[e.g.,][]{Hernandez07,Cervantes-Sodi08}. Thus, while imperfect for individual cases, our estimator serves as a statistically meaningful proxy for halo spin. Crucially, random errors introduced by these approximations primarily increase scatter and thereby tend to dilute\---not artificially generate\---intrinsic correlations.

Finally, we fully acknowledge that studying the relationship between halo spin and $d_{\rm gf}$ (or filamentary environmental density) using unresolved HI data and semi-analytical relations is inherently approximate and subject to significant uncertainties. Nevertheless, single-dish surveys like ALFALFA offer unparalleled statistical power due to their large samples, whereas current interferometric surveys remain limited by smaller number statistics. Consequently, our work is designed as a population-level statistical study aimed at detecting coherent environmental trends rather than delivering precise measurements for individual galaxies. In future work, we plan to leverage spatially resolved HI kinematics from interferometers combined with large-volume cosmological hydrodynamical simulations to investigate this connection more robustly. Such an approach will substantially improve reliability and enable direct validation of the current results. However, this lies beyond the scope of the present study.
\\


\acknowledgments

{\bf We thank the referee for providing constructive feedback.} Y.R. acknowledges supports from the CAS Pioneer Hundred Talents Program (Category B), NSFC grants 12273037 and 12522302, and the USTC Research Funds of the Double First-Class Initiative. This work is supported by the China Manned Space Program with grant no. CMS-CSST-2025-A06 and CMS-CSST-2025-A08.





\end{document}